# Ultrafast controlling net magnetization in *g*-wave altermagnets *via* laser fields

Zhaobo Zhou,[1] Sangeeta Sharma,[2,3,] and Junjie He[1,]

[1]Faculty of Science, Charles University, Prague 12843, Czech Republic
[2]Max-Born-Institute for Non-linear Optics and Short Pulse Spectroscopy, 12489 Berlin, Germany
[3]Institute for Theoretical Solid-State Physics, Freie Universität Berlin, 14195 Berlin, Germany

The diverse nodal spin structures in d/g/i-wave altermagnets (AM) may cause distinct light-induced spin responses yet remain poorly understood. Using time-dependent density functional theory (TDDFT), we reveal that laser-induced ultrafast demagnetization dynamics in the *g*-wave AM CrSb are strongly governed by the laser incidence direction. Under normal incidence along the [0001] axis, two Cr sublattices exhibit symmetric temporal demagnetization but with different amplitudes, preserving the net-zero magnetization—unlike the behavior in *d*-wave AM. Off-normal incidence, however, induces pronounced asymmetric demagnetization between sublattices, transiently driving the system into a ferrimagnetic-like state with a sizable net magnetization. This direction-dependent response arises from the characteristic nodal structures in bulk *g*-wave AM's electronic structure, which enable anisotropic optical intersite spin transfer (OISTR). By comparing *g*-wave and *d*-wave AMs, we propose that light-induced magnetization arises when laser polarization aligns with spin-uncompensated regions in electronic structures. This can be readily determined from the local spin density of states along specific band paths. Our results provide a fundamental understanding for laser-induced ultrafast dynamics in AM.

Keywords: *g*-wave altermagnets, CrSb, ultrafast spin dynamics, demagnetization, TDDFT

## INTRODUCTION

Magnetism lies at the heart of condensed matter physics and modern spintronics. For decades, magnetic materials have been classified into two distinct phases: ferromagnets (FM) which exhibit net magnetization and broken time-reversal symmetry, and antiferromagnets (AFMs), where collinear spin ordering cancels out the macroscopic magnetization. However, this conventional paradigm has recently been upended by the discovery of *altermagnets* (AM)—a fundamentally new class of magnetic order that defies conventional classification (*1*, *2*). AM combines two seemingly incompatible features: time-reversal symmetry breaking, like in FM, and zero net magnetization, as in AFM (*3*, *4*). These systems exhibit alternating spin-splitting in momentum space even without spin-orbit coupling (SOC) in a compensated zero-magnetization, as well as anisotropic, even-parity spin-split isoenergy surfaces with *d*-, *g*-, or *i*-wave symmetry (*5*, *6*).

The symmetry-protected spin splitting and anisotropic band structures of AMs give rise to unconventional charge, spin, and heat transport (*7–9*), enabling exotic current-driven spin phenomena beyond conventional magnets. These distinctive features position AMs as a compelling platform for next-generation spintronic applications (*10*). Beyond their transport phenomena, the unique nodal electronic structure of AMs also opens exciting possibilities for unconventional ultrafast spin responses under laser excitation.

Since the pioneering observation of femtosecond demagnetization in ferromagnetic nickel film (*11*), the field of ultrafast magnetism has largely focused on FMs and AFMs, leaving laser-driven magnetization dynamics in AMs largely unexplored. A key theoretical breakthrough in this regard was the optical-induced intersite spin transfer (OISTR) effect (*12*), which showed that optical excitation can coherently redistribute spins between magnetic sublattices in multicomponent magnetic systems (*13–17*). While initial theoretical efforts have begun to explore optical signatures and spin polarization in AMs (*18–21*), experimental investigations remain limited. Some studies have probed AMs under laser excitation using techniques such as time-resolved magneto-optical Kerr effect, angle-resolved photoemission spectroscopy, and magnetic circular dichroism (*10*, *22–27*).

Our previous theoretical work revealed distinct ultrafast magnetization dynamics in *d*-wave AM compared to conventional magnets (*21*). However, bulk *g*-wave AM possesses a qualitatively different spin texture, featuring both vertical and horizontal nodal planes in momentum space —suggesting fundamentally different pathways for electron and spin dynamics under ultrafast laser excitation. Despite these intriguing possibilities, a systematic investigation of laser-induced spin dynamics and the underlying microscopic mechanisms in bulk *g*-wave AMs remains largely unexplored.

In this work, we systematically investigate the laser-induced femtosecond spin dynamics in the bulk *g*-wave AM CrSb using real-time time-dependent density functional theory (TDDFT). We find that linearly polarized laser pulses with varying incidence angles can induce either symmetric or asymmetric demagnetization of the $Cr_1$ and $Cr_2$ sublattices, where the asymmetric spin response may further give rise to ferrimagnetic-like state with a metastable magnetization. These direction-dependent spin dynamics originate from the nodal structures inherent to the *g*-wave electronic structure, which result in a highly anisotropic OISTR effect. We further compare the spin dynamics in *g*-wave and *d*-wave AM, highlighting key differences in their spin dynamics and the emergence of net magnetization. Based on these insights, we propose a practical criterion—derived from the local spin density of states along specific band paths—that can guide the tuning of laser parameters to induce magnetization in otherwise compensated AM.

## RESULTS

***Symmetric LDOS along specific band paths:*** CrSb crystallizes in a hexagonal lattice and exhibits a collinear AFM state, with the Néel vector aligned antiparallel along the *c*-axis as shown in Fig. 1A. (Hereafter, laser incidence along this axis is referred to as *normal incidence*) The spin-resolved band structures without SOC exhibit strong momentum space-dependent behavior: complete spin degeneracy along the K-Γ-K′ path (Fig. 1C), while pronounced spin splitting along the S′-A′-U′ or S-A-U paths (Fig. 1D), leading to alternating, yet fully compensated spin polarization in momentum space —a fingerprint of bulk g-wave AM (Fig. 1B). We further analyze the sublattice-resolved spin-polarized local density of states (LDOS) for Cr1 and Cr2 atoms along specific band paths, including K-Γ, Γ-K′, S′-A′ (A-U) and A′-U′ (S-A) paths, see Fig. 1 E to H. We can see that the majority state of Cr1 and Cr2 below the Fermi level along these paths is contributed from the spin-up state and spin-down state, respectively, explicating the antiparallel magnetic order of the two Cr sublattices.

In conventional AFM, the LDOS of spin-up and spin-down sublattices are symmetrically distributed, leading to fully compensated local moments and zero net magnetization. In contrast, the g-wave AM CrSb exhibits fundamentally different LDOS behavior, stemming from its distinctive vertical and horizontal nodal planes—markedly different from both traditional AFM and d-wave AM (21). At spin-degenerate nodal planes (K-Γ-K′ path), the LDOS between sublattices remains fully compensated, consistent with AFM-like symmetry (Fig. 1F and 1H). However, despite opposite spin polarization and unequal LDOS amplitudes along the S′–A′ and A′–U′ segments, their overall contributions are symmetric, resulting in complete compensation along the full S′–U′ path. As a result, theoretically, under normal laser incidence—where the polarization lies within the ab-plane and equally spans both segments (S′–A′ and A′–U′)—no net spin transfer occurs between sublattices, and the system exhibits symmetric laser-induced demagnetization without generating net magnetization. Nevertheless, the magnitude of demagnetization with various laser polarizations can still differ, due to the unequal LDOS amplitudes in different momentum regions.

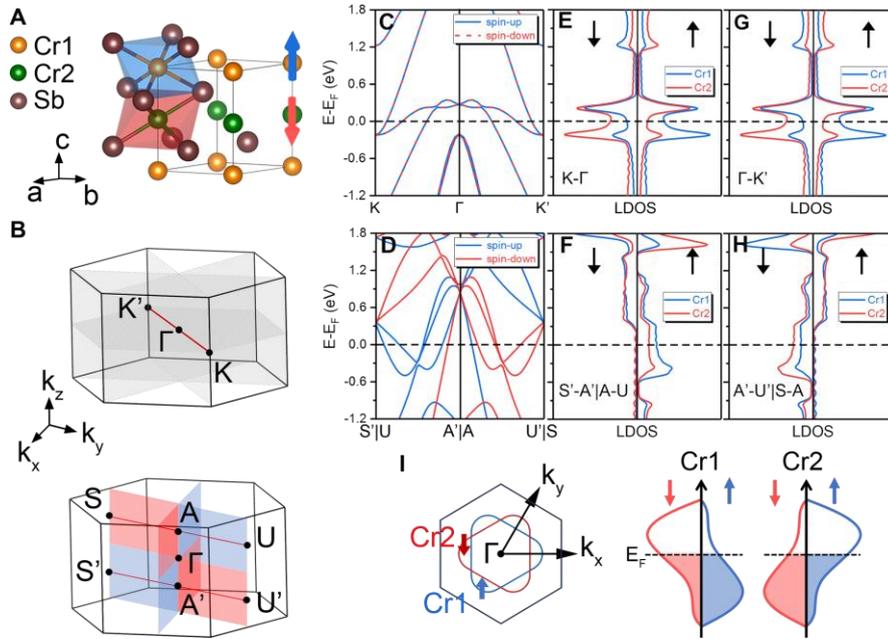

**Fig. 1. Ground-state geometry and electronic structure of CrSb.** (**A**) Optimized crystal structure of CrSb with two Cr sublattices. Yellow, green and brown spheres represent Cr1, Cr2 and Sb atoms, respectively. Blue and red arrows represent opposite Néel vectors of Cr atoms. (**B**) 3D Brillouin zone of CrSb showing four spin-degenerate nodal planes (gray) and two spin-polarized planes. Blue and red colors highlight the spin-up and spin-down polarization respectively, showing a compensated behavior along the S-A-U (S′-A′-U′) path. (**C**) and (**d**) Band structures of CrSb without SOC along the spin-degenerate K-Γ-K′ and spin-polarized S-A-U (S′-A′-U′) paths, respectively. (**D** to **H**) Spin-resolved LDOS of Cr1 and Cr2 atoms along the K-Γ, Γ-K′, S′-A′ (A-U) and A′-U′ (S-A) paths, respectively. (**I**) Schematic constant-energy contour at the $k_z=-0.25$ Å$^{-1}$ plane (left panel) and spin-resolved LDOS of two Cr atoms (right panel). Blue and red donate the spin-up and spin-down states, respectively.

***Symmetric demagnetization with direction-dependent amplitude (Normal incidence):*** To verify the hypothesis derived from LDOS analysis, we perform real-time TDDFT simulations to investigate the ultrafast demagnetization dynamics induced by linearly polarized laser pulses along normal incidence, focusing on how varying the laser polarization angle $\theta$ controls the spin moment response of Cr sublattices. Here, $\theta$ donates the in-plane (a-b plane) polarization angle between the electric field vector $E_\theta$ of the laser pulse and the $k_x$ axis within the $k_x$–$k_y$ plane, see Fig. 2A. We select two representative directions: (I) $\theta=0°/60°$, corresponding to $E_\theta$ parallel to the spin-polarized planes; and (II) $\theta=30°/90°$, corresponding to $E_\theta$ parallel to the spin-degenerate nodal planes. The time evolution of atom-resolved spin moments reveals a remarkable angle-dependence, in which two Cr atoms exhibit stronger demagnetization at $\theta=30°/90°$ than that at $\theta=0°/60°$ (Fig. 2B). Importantly, the two Cr sublattices respond identically across all angles, resulting in symmetric demagnetization and a vanishing net magnetization—consistent with the compensated nature of the AFM. In contrast, this is fundamentally different from $d$-wave AM, where the emergence of magnetization strongly depends on its distinct nodal structures (*21*). These spin dynamics in CrSb are in excellent agreement with the predictions based on LDOS analysis.

To further understand the $\theta$-dependent symmetric demagnetization of two Cr sublattices, we analyze their transient spin-resolved LDOS at t=32.5 fs (Fig. 2D). The higher transient occupation of minority states in Cr atoms at $\theta=0°/60°$ suggests a more efficient OISTR effect, leading to stronger demagnetization compared to $\theta=30°/90°$. Besides, Cr1 and Cr2 exhibit identical LDOS evolution for a specific $\theta$: spin-up states of Cr2 and spin-down states of Cr1 increase equally with the time evolution (Fig. 2C). This identical change in LDOS further explains why the symmetric demagnetization between two Cr sublattices is independent of the laser incidence angle $\theta$.

Based on the above analysis, we propose the physical picture of $\theta$-dependent symmetric demagnetization behavior in CrSb, as illustrated in Fig. 2E: under laser excitation, the OISTR process between the Cr sublattices remains symmetric and independent of the polarization angle $\theta$, owing to laser-induced identical charge distribution between the two Cr sublattices, resulting in zero net magnetic moment between the sublattices at any $\theta$. Notably, the magnitude of the OISTR effect exhibits a strong dependence on $\theta$, being significantly enhanced when the polarization direction is oriented with the compensated spin-polarized planes. This leads to a more pronounced demagnetization in the Cr sublattices. This can also be understood by variation in LDOS amplitude along the specific band paths. Notably, this physical picture is fundamentally different from that with $d$-wave AM, where normal incidence induces asymmetric spin dynamics and net magnetization, such dynamics will be controllable via polarization angle $\theta$ (*21*).

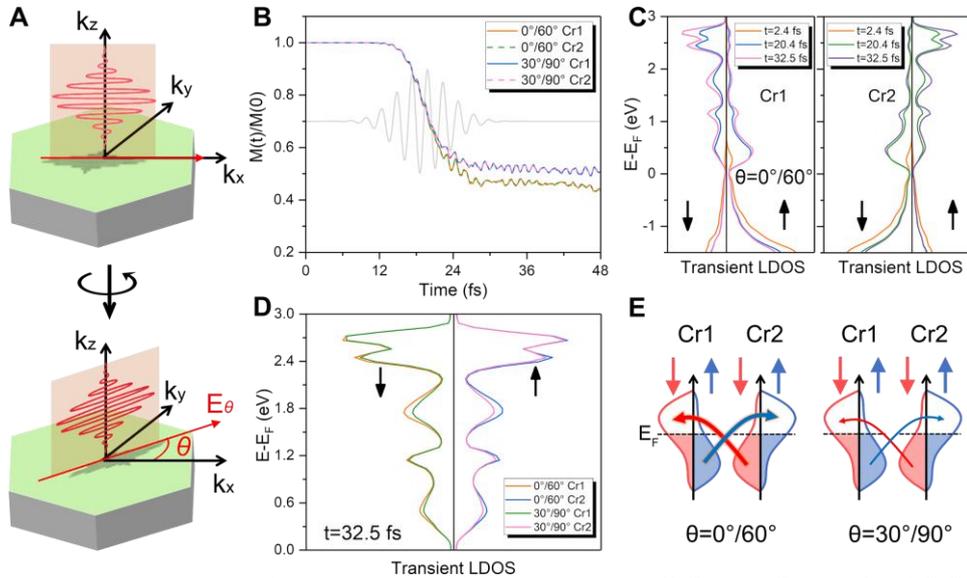

**Fig. 2. Laser-induced symmetric demagnetization dynamics in CrSb.** (**A**) Schematic of CrSb under laser pulse irradiation with in-plane excitation angle $\theta$, which is defined as the angle between the electric field vector $E_\theta$ of the laser and the $k_x$ axis in the $k_x$–$k_z$ plane. (**B**) Normalized spin moment loss of Cr1 and Cr2 as a function of time at $\theta=0°$, 30°, 60° and 90°. The vector potential of the laser pulse is shown in gray (central frequency 1.63 eV, full width at half maximum (FWHM) of ~10 fs, and an incident fluence of 9.8 mJ/cm$^2$). (**C**) Transient populated LDOS of the Cr1 and Cr2 with $\theta=0°/60°$ at t=2.4 fs, 20.4 fs and 32.5 fs, respectively. Up- and down-pointing arrows mark the spin-up and spin-down states, respectively. (**D**) Transiently populated part of LDOS at t=32.5 fs with $\theta=0°/60°$ and $\theta=30°/90°$. (**E**) Schematic of symmetric

OISTR process between Cr1 and Cr2 atoms with $\theta=0°/60°$ and $\theta=30°/90°$. Blue and red arrows represent the spin-up and spin-down states, respectively.

*Asymmetrical demagnetization induced by off-normal incidence:* So far, we have revealed the laser-induced symmetric demagnetization between the Cr sublattices along the direction parallel to the $k_z$ axis in the BZ. To further explore its angular dependence beyond angle $\theta$, we introduce an additional off-plane polarization angle $\varphi$ by tilting the laser incidence away from the $k_z$ axis (Hereafter, referred to as *off-normal incidence*), see Fig. 3A. Upon off-normal incidence at specific $\varphi=\pm56°$, the spin moment loss of Cr1 and Cr2 becomes asymmetric after 15 fs, leading to a finite net magnetic moment ($\Delta M\neq 0$), see Fig. 3 (B and D). Strikingly, $\Delta M$ is highly sensitive to both $\varphi$ and $\theta$: For $\theta=0°/60°$, even slight deviations from $\varphi=0°$ to $\pm90°$ induce asymmetric demagnetization between Cr1 and Cr2 (fig. S2). In contrast, for $\theta=30°/90°$, the demagnetization remains symmetric between the Cr atoms (i.e., $\Delta M=0$) regardless of the $\varphi$ (fig. S3). Consequently, we map the $\Delta M$ between the Cr sublattice as a function of both $\theta$ and $\varphi$, see Fig. 3C. The non-zero $\Delta M$ appears in four selected angular regions, defined by combinations of $\theta$ and $\varphi$: (I) $0°<\theta<30°$ and $-90°<\varphi<0°$; (II) $0°<\theta<30°$ and $0°<\varphi<90°$; (III) $30°<\theta<90°$ and $-90°<\varphi<0°$; (IV) $30°<\theta<90°$ and $0°<\varphi<90°$. These results highlight that both the in-plane and out-of-plane components of the laser incidence critically dominate the activation of spin-polarized transitions, thus enriching the degrees of freedom for controlling magnetization dynamics.

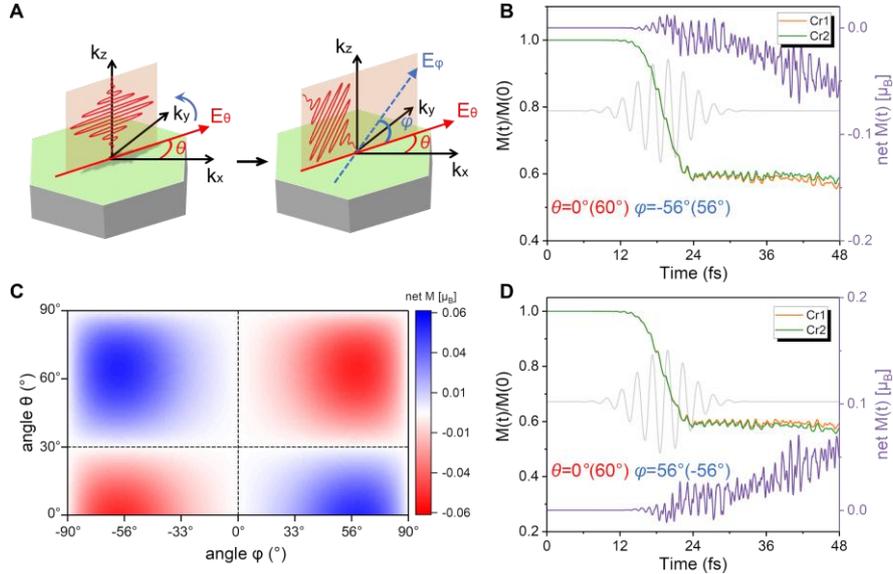

**Fig. 3. Laser-induced asymmetric demagnetization dynamics in CrSb.** (**A**) Schematic of CrSb under laser pulse excitation with in-plane angle $\theta$ and off-plane ($k_z$–$E_\theta$ plane) angle $\varphi$. The dotted blue arrow marks the spatial electric field vector $E_\varphi$ of the laser. (**B** and **D**) Normalized spin moment loss of Cr1 and Cr2 as a function of time at $\theta=0°$ $\varphi=-56°$ ($\theta=60°$ $\varphi=56°$) and $\theta=0°$ $\varphi=56°$ ($\theta=60°$ $\varphi=-56°$), respectively. The net magnetic moment (net M) is shown in purple. (**C**) The net M for all combinations of angle $\theta$ and $\varphi$.

*Asymmetrical LDOS and uncompensated spin-polarization in off-normal direction:* To gain deeper insights into the microscopic mechanism responsible for the $\varphi$-induced symmetry breaking in demagnetization, we first calculate and analyze the static spin-resolved LDOS of Cr atoms along the specific R-Γ-R′ band path that is parallel to the off-plane electric field vector $E\varphi$ of the laser, see Fig. 4A. For both Cr1 and Cr2, the electronic states near the Fermi level exhibit asymmetrical with net spin polarization (uncompensated spin) that primarily arises from spin-down components, in contrast to the compensated behavior observed along the S′-A′-U′ path (fig. S1). This indicates a transition of spin nodal structures from symmetric *g*-wave to asymmetric *d*-wave (Fig. 4C) in terms of the topological structures of spin distribution (also see Fig. 5), thereby enabling the laser to induce asymmetric spin transfer once the laser is incident with a finite off-plane polarization angle $\varphi$. This arises from the coexistence of vertical and horizontal nodal planes with differing spin polarizations in momentum space in bulk *g*-wave

AM.

To substantiate this, we further calculate the transient spin-resolved LDOS of minority states for both Cr atoms at t=45.7 fs, as depicted in Fig. 4B. It is clear that the transient occupation of minority states in Cr2 has a larger integrated value compared to that in Cr1, indicating an unequal OISTR between the two Cr sublattices. These results provide a microscopic picture for the *off-normal incidence* induced asymmetric demagnetization and net magnetization between Cr sublattices, as illustrated in Fig. 4D. Upon off-normal incidence at $0°<\varphi<90°$, more spin-up electrons are excited into the minority states of Cr2 than those of Cr1, leading to a spin-selective charge redistribution between the two Cr sublattices and asymmetric demagnetization consequently. Vice versa, the opposite OISTR process can also be observed in the range of $-90°<\varphi<0°$.

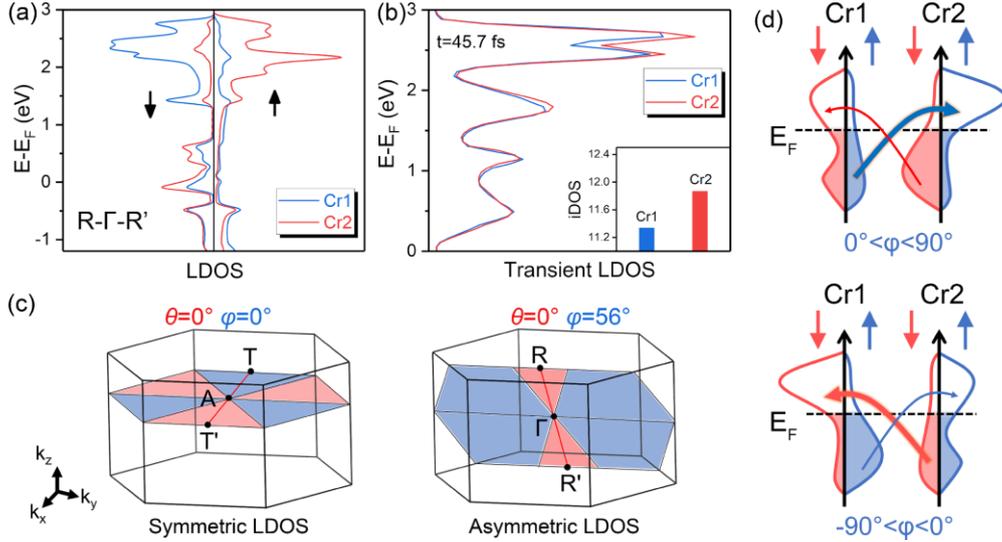

**Fig. 4. Out-of-plane spin-polarized excitation in CrSb.** (**A**) Spin-resolved LDOS of two Cr sublattices along the R-Γ-R' path. (**B**) Transiently populated minority states in the LDOS of Cr1 and Cr2 at t=45.7 fs. The inset plane shows the corresponding integrated value of LDOS (iDOS) for Cr1 and Cr2. (**C**) Schematic of spin structure of the electronic state near the Fermi level for two planes in the BZ. Blue and red colors highlight the spin-up and spin-down polarization, respectively. Red lines represent the direction of the electric field vector of the laser at $\varphi=0°$ and $56°$, respectively. (**D**) Schematic of unequal OISTR between two Cr sublattices at $0°<\varphi<90°$ and $-90°<\varphi<0°$.

Previous studies have established that AM exhibits pronounced spin splitting even in the absence of spin-orbit coupling (SOC), with only a minor additional splitting when SOC is included. To further assess the role of SOC in the laser-driven demagnetization process, we compute and compare the spin moment loss of Cr atoms with and without SOC, as shown in fig. S4. Remarkably, our results reveal that the symmetric demagnetization between two Cr atoms persists regardless of SOC inclusion. Specifically, without SOC, the two Cr atoms undergo the equivalent spin moment for both $\theta=0°/60°$ and $\theta=30°/90°$. However, when SOC is introduced, the demagnetization of two Cr atoms is amplified, while it retains its symmetry. We attribute this enhancement to SOC-driven spin splitting, which strengthens the spin current into unoccupied minority states of the opposite sublattice, effectively intensifying the spin transfer process. As a result, SOC intensifies demagnetization and highlights its essential role in modulating spin dynamics in CrSb.

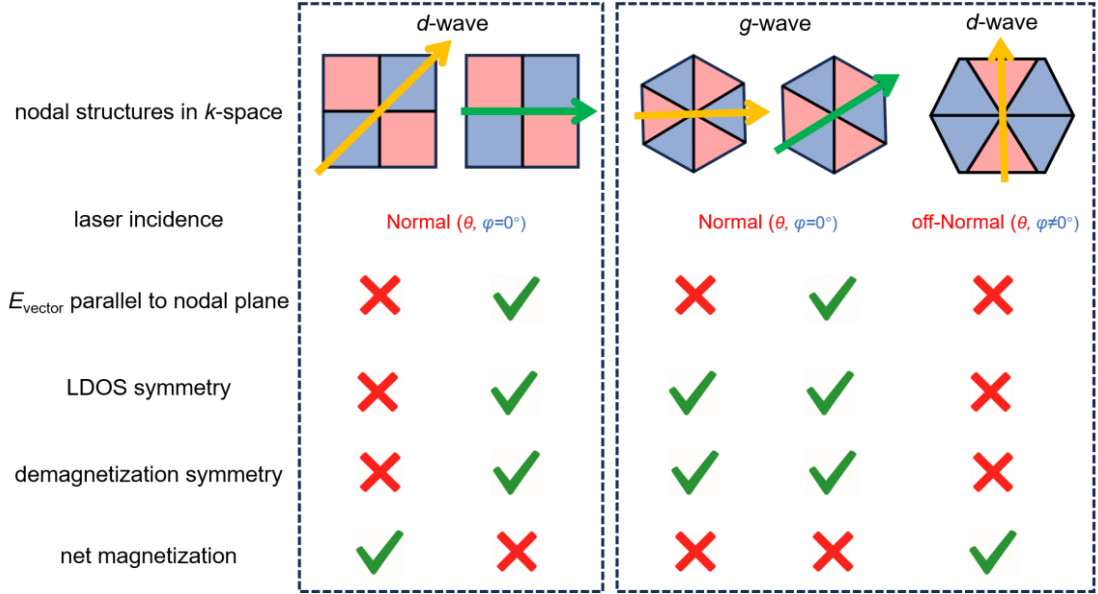

**Fig. 5. Illustration of laser-driven asymmetric/symmetric demagnetization in *d/g*-wave AMs.** Red and blue regions represent two spin-polarized electronic structures of the valence band maximum in the BZ, respectively. Yellow and green arrows represent the laser polarization direction (electric field vector $E_{vector}$) parallel to the spin-polarized nodal planes and spin-degenerate nodal planes, respectively.

**DISCUSSION**

Our previous work demonstrated the laser-induced asymmetric demagnetization of the Ru sublattices in planar *d*-wave AM $RuO_2$ (*21*). Building upon this work, here we systematically compare the laser-induced demagnetization dynamics, net magnetization emergence, nodal structures and underlying physical mechanism for *d/g*-wave AM as summarized in Fig. 5. We further propose a general criterion to determine the occurrence of asymmetric spin dynamics and magnetization in AM: when laser polarization (electric field vector $E_{vector}$) aligns with spin-uncompensated regions (highlighted in blue and red in Fig. 5) within electronic structure. These spin-resolved regions (SRR) can be readily determined from the LDOS of sublattice along specific band paths.

Specifically, when the laser polarization coincides with a symmetrical SRR or LDOS, the symmetric demagnetization occurs across the magnetic sublattice. In contrast, laser incidence aligned with an asymmetrical SRR or LDOS path leads to an asymmetric demagnetization and a nonzero net magnetization between the magnetic sublattices. For instance, in CrSb, the SRR shows a compensated alternating pattern at $\theta=0°$ and $30°$, while at $\varphi \neq 0°$, this SRR becomes uncompensated (Fig. 4C and fig. S5), which is similar to that in planar *d*-wave AM.

Notably, for $\theta=30°$ and $\varphi \neq 0°$, the demagnetization remains symmetric as the polarization direction lies within the spin-degenerate nodal plane (fig. S5). Nevertheless, according to our proposed criterion, further rotating the laser by an additional angle $\alpha$ to polarization direction within a spin-uncompensated SRR is expected to induce asymmetric demagnetization (fig. S5, C and D).

This universal criterion provides a robust and accessible strategy— relying solely on *ab initio* ground-state calculations — to guide optical control of magnetization dynamics in AM. based on only ab initio calculations, for the optical control of the magnetization dynamics in AMs. Overall, the LDOS of sublattice along specific band paths offers a straightforward way to predict and manipulate spin dynamics and net magnetization in these systems.

In summary, we have unveiled the ultrafast femtosecond demagnetization dynamics in bulk *g*-wave AM CrSb through state-of-the-art real-time ab initio simulations, revealing a striking dependence on laser incidence direction. Specifically, normal incidence triggers a perfectly symmetric demagnetization between the two Cr sublattices, regardless of the in-plane polarization angle $\theta$, while off-normal incidence breaks this symmetry, driving a ferrimagnetic-like phase transition accompanied by a substantial net magnetization. This rich direction-dependent behavior stems directly from the unique nodal architecture of bulk *g*-wave AM— where vertical and horizontal nodal planes carve out momentum-space regions with opposite spin polarizations. Such exotic electronic structures lead to an anisotropic OISTR effect, enabling highly spin-selective charge redistribution between sublattices under laser excitation. Beyond these insights, we propose a universal and practical criterion for laser control of magnetization dynamics in AM: the emergence of asymmetric spin behavior is dictated by whether the laser polarization aligns with spin-uncompensated regions. This region can be readily determined from the local spin density of states of sublattices along specific band paths. Our work not only uncovers the microscopic mechanisms driving laser-induced spin dynamics in *g*-wave AM but also lays out a simple yet powerful strategy to predict and tailor ultrafast magnetization responses across a broad family of these emerging AMs.

**METHODS**
To investigate the laser-induced demagnetization dynamics in CrSb, we employed state-of-the-art real-time time-dependent density functional theory (rt-TDDFT) methods (*28*). This approach systematically transforms the computational challenge of electron interactions into solving the Kohn-Sham (KS) equation for noninteracting Fermions within an artificial potential. The time-dependent KS equation is:

$$i\frac{\partial \psi_j(r,t)}{\partial t} = \left[\frac{1}{2}\left(-i\nabla + \frac{1}{c}(A(t) + A_{XC}(t))\right)^2 + v_s(r,t)\right]\psi_j(r,t) \quad (1)$$

Where $\psi_j$ is a KS orbital and the KS effective potential $v_s(r,t) = v(r,t) + v_H(r,t) + v_{xc}(r,t)$ consists of the external potential $v$, the classical Hartree potential $v_H$, and the exchange-correlation (XC) potential $v_{xc}$, respectively. The vector potential $A(t)$ represents the applied laser pulse within the dipole approximation and $A_{XC}(t)$ the XC vector potential.

The structural optimization was implemented with the VASP (*29, 30*). The Perdew-Burke-Ernzerhof (PBE) functional within the generalized gradient approximation (GGA) was employed for exchange-correlation interactions (*31*). Electron-ion interaction was described using the projector-augmented wave method (*32*). A cutoff energy of 500 eV and a Monkhorst-Pack 9×9×6 k-mesh grid were utilized. The lattice constants and atomic positions were fully relaxed until the atomic forces were smaller than 0.1 meV Å$^{-1}$. The electron relaxation convergence criterion was 10$^{-7}$ eV. Hubbard corrections were applied to account for the strongly correlated *d*-electrons of Cr with $U_{eff}$ = 3 eV.

Laser-induced spin dynamics simulations were implemented with the ELK code (*33*) using a fully noncollinear version of rt-TDDFT. The calculations employed a 6·6·8 *k*-point grid, a smearing width of 0.027 eV, and a time step of Δt = 0.1 atomic unit (a.u.). All calculations were performed using the adiabatic local spin density approximation (ALSDA).

**SUPPLEMENTARY MATERIALS**
This PDF file includes:
Figs. S1 to S5

**Acknowledgements:** We thank the e-INFRA CZ (ID:90140) for providing computational resources. **Funding:** Z. Z would like to thank the support from MSCA Fellowships CZ-UK3 CZ.02.01.01/00/22_010/0008820. S. S would like to thank DFG for funding through project-ID 328545488 TRR227 (project A04) and Leibniz Professorin program (SAWP118/2021). **Author contributions:** Z. Z and J. H designed the project. Z. Z performed the calculation. S. S contributed to the analysis. All authors contributed to the writing of the manuscript. **Competing interests:** The authors declare that they have no competing interests. **Data and materials availability:** All data needed to evaluate the conclusions in the paper are present in the paper and/or the Supplementary Materials. The code used in the manuscript is accessible at https://elk.sourceforge.io/.

**REFERENCES AND NOTES**
1. L. Šmejkal, J. Sinova, T. Jungwirth, Emerging Research Landscape of Altermagnetism.

*Phys. Rev. X* **12**, 040501 (2022).

2. L. Šmejkal, J. Sinova, T. Jungwirth, Beyond Conventional Ferromagnetism and Antiferromagnetism: A Phase with Nonrelativistic Spin and Crystal Rotation Symmetry. *Phys. Rev. X* **12**, 031042 (2022).

3. L. Šmejkal, R. González-Hernández, T. Jungwirth, J. Sinova, Crystal time-reversal symmetry breaking and spontaneous Hall effect in collinear antiferromagnets. *Sci. Adv.* **6**, eaaz8809 (2020).

4. O. Fedchenko, J. Minár, A. Akashdeep, S. W. D'Souza, D. Vasilyev, O. Tkach, L. Odenbreit, Q. Nguyen, D. Kutnyakhov, N. Wind, L. Wenthaus, M. Scholz, K. Rossnagel, M. Hoesch, M. Aeschlimann, B. Stadtmüller, M. Kläui, G. Schönhense, T. Jungwirth, A. B. Hellenes, G. Jakob, L. Šmejkal, J. Sinova, H.-J. Elmers, Observation of time-reversal symmetry breaking in the band structure of altermagnetic RuO2. *Sci. Adv.* **10**, eadj4883 (2024).

5. L. Šmejkal, A. H. MacDonald, J. Sinova, S. Nakatsuji, T. Jungwirth, Anomalous Hall antiferromagnets. *Nat. Rev. Mater.* **7**, 482–496 (2022).

6. R. Zarzuela, R. Jaeschke-Ubiergo, O. Gomonay, L. Šmejkal, J. Sinova, Transport theory and spin-transfer physics in d-wave altermagnets. *Phys. Rev. B* **111** (2025).

7. H. Bai, L. Han, X. Y. Feng, Y. J. Zhou, R. X. Su, Q. Wang, L. Y. Liao, W. X. Zhu, X. Z. Chen, F. Pan, X. L. Fan, C. Song, Observation of Spin Splitting Torque in a Collinear Antiferromagnet RuO 2. *Phys. Rev. Lett.* **128**, 197202 (2022).

8. R. D. Gonzalez Betancourt, J. Zubáč, R. Gonzalez-Hernandez, K. Geishendorf, Z. Šobáň, G. Springholz, K. Olejník, L. Šmejkal, J. Sinova, T. Jungwirth, S. T. B. Goennenwein, A. Thomas, H. Reichlová, J. Železný, D. Kriegner, Spontaneous Anomalous Hall Effect Arising from an Unconventional Compensated Magnetic Phase in a Semiconductor. *Phys. Rev. Lett.* **130**, 036702 (2023).

9. H.-Y. Ma, M. Hu, N. Li, J. Liu, W. Yao, J.-F. Jia, J. Liu, Multifunctional antiferromagnetic materials with giant piezomagnetism and noncollinear spin current. *Nat. Commun.* **12**, 2846 (2021).

10. O. J. Amin, A. Dal Din, E. Golias, Y. Niu, A. Zakharov, S. C. Fromage, C. J. B. Fields, S. L. Heywood, R. B. Cousins, F. Maccherozzi, J. Krempaský, J. H. Dil, D. Kriegner, B. Kiraly, R. P. Campion, A. W. Rushforth, K. W. Edmonds, S. S. Dhesi, L. Šmejkal, T. Jungwirth, P. Wadley, Nanoscale imaging and control of altermagnetism in MnTe. *Nature* **636**, 348–353 (2024).

11. E. Beaurepaire, J.-C. Merle, A. Daunois, J.-Y. Bigot, Ultrafast Spin Dynamics in Ferromagnetic Nickel. *Phys. Rev. Lett.* **76**, 4250–4253 (1996).

12. J. K. Dewhurst, P. Elliott, S. Shallcross, E. K. U. Gross, S. Sharma, Laser-Induced Intersite Spin Transfer. *Nano Lett.* **18**, 1842–1848 (2018).

13. S. A. Ryan, P. C. Johnsen, M. F. Elhanoty, A. Grafov, N. Li, A. Delin, A. Markou, E. Lesne, C. Felser, O. Eriksson, H. C. Kapteyn, O. Granas, M. M. Murnane, Optically controlling the competition between spin flips and intersite spin transfer in a Heusler half-metal on sub–100-fs time scales. *Sci. Adv.* **9**, eadi1428 (2023).

14. E. Golias, I. Kumberg, I. Gelen, S. Thakur, J. Gordes, R. Hosseinifar, Q. Guillet, J. K.

Dewhurst, S. Sharma, C. Schussler-Langeheine, N. Pontius, W. Kuch, Ultrafast Optically Induced Ferromagnetic State in an Elemental Antiferromagnet. *Phys. Rev. Lett.* **126**, 107202 (2021).

15. C. Möller, H. Probst, G. S. M. Jansen, M. Schumacher, M. Brede, J. K. Dewhurst, M. Reutzel, D. Steil, S. Sharma, S. Mathias, Verification of ultrafast spin transfer effects in iron-nickel alloys. *Commun. Phys.* **7**, 1–6 (2024).

16. F. Siegrist, J. A. Gessner, M. Ossiander, C. Denker, Y.-P. Chang, M. C. Schröder, A. Guggenmos, Y. Cui, J. Walowski, U. Martens, J. K. Dewhurst, U. Kleineberg, M. Münzenberg, S. Sharma, M. Schultze, Light-wave dynamic control of magnetism. *Nature* **571**, 240–244 (2019).

17. W.-B. Lee, S. Hwang, H.-W. Ko, B.-G. Park, K.-J. Lee, G.-M. Choi, Spin-torque-driven gigahertz magnetization dynamics in the non-collinear antiferromagnet Mn3Sn. *Nat. Nanotechnol.* **20**, 487–493 (2025).

18. M. Weber, S. Wust, L. Haag, A. Akashdeep, K. Leckron, C. Schmitt, R. Ramos, T. Kikkawa, E. Saitoh, M. Kläui, L. Šmejkal, J. Sinova, M. Aeschlimann, G. Jakob, B. Stadtmüller, H. C. Schneider, All optical excitation of spin polarization in d-wave altermagnets. arXiv arXiv:2408.05187 [Preprint] (2024). http://arxiv.org/abs/2408.05187.

19. M. Weber, K. Leckron, L. Haag, R. Jaeschke-Ubiergo, L. Šmejkal, J. Sinova, H. C. Schneider, Ultrafast electron dynamics in altermagnetic materials. arXiv arXiv:2411.08160 [Preprint] (2024). https://doi.org/10.48550/arXiv.2411.08160.

20. L. Haag, M. Weber, K. Leckron, L. Šmejkal, J. Sinova, H. C. Schneider, Optical signatures of bulk g-wave altermagnetism in MnTe. arXiv arXiv:2505.13415 [Preprint] (2025). https://doi.org/10.48550/arXiv.2505.13415.

21. Z. Zhou, S. Sharma, J. K. Dewhurst, J. He, Magnetizing altermagnets by ultrafast asymmetric spin dynamics. arXiv arXiv:2502.01258 [Preprint] (2025). https://doi.org/10.48550/arXiv.2502.01258.

22. J. Ding, Z. Jiang, X. Chen, Z. Tao, Z. Liu, T. Li, J. Liu, J. Sun, J. Cheng, J. Liu, Y. Yang, R. Zhang, L. Deng, W. Jing, Y. Huang, Y. Shi, M. Ye, S. Qiao, Y. Wang, Y. Guo, D. Feng, D. Shen, Large Band Splitting in g-Wave Altermagnet CrSb. *Phys. Rev. Lett.* **133**, 206401 (2024).

23. J. Krempaský, L. Šmejkal, S. W. D'Souza, M. Hajlaoui, G. Springholz, K. Uhlířová, F. Alarab, P. C. Constantinou, V. Strocov, D. Usanov, W. R. Pudelko, R. González-Hernández, A. Birk Hellenes, Z. Jansa, H. Reichlová, Z. Šobáň, R. D. Gonzalez Betancourt, P. Wadley, J. Sinova, D. Kriegner, J. Minár, J. H. Dil, T. Jungwirth, Altermagnetic lifting of Kramers spin degeneracy. *Nature* **626**, 517–522 (2024).

24. T. Osumi, S. Souma, T. Aoyama, K. Yamauchi, A. Honma, K. Nakayama, T. Takahashi, K. Ohgushi, T. Sato, Observation of a giant band splitting in altermagnetic MnTe. *Phys. Rev. B* **109**, 115102 (2024).

25. A. Hariki, A. Dal Din, O. J. Amin, T. Yamaguchi, A. Badura, D. Kriegner, K. W. Edmonds, R. P. Campion, P. Wadley, D. Backes, L. S. I. Veiga, S. S. Dhesi, G. Springholz, L. Šmejkal, K. Výborný, T. Jungwirth, J. Kuneš, X-Ray Magnetic Circular Dichroism in Altermagnetic α-MnTe. *Phys. Rev. Lett.* **132**, 176701 (2024).

26. M. Hajlaoui, S. Wilfred D'Souza, L. Šmejkal, D. Kriegner, G. Krizman, T. Zakusylo, N.

Olszowska, O. Caha, J. Michalička, J. Sánchez-Barriga, A. Marmodoro, K. Výborný, A. Ernst, M. Cinchetti, J. Minar, T. Jungwirth, G. Springholz, Temperature Dependence of Relativistic Valence Band Splitting Induced by an Altermagnetic Phase Transition. *Adv. Mater.* **36**, 2314076 (2024).

27. Y.-P. Zhu, X. Chen, X.-R. Liu, Y. Liu, P. Liu, H. Zha, G. Qu, C. Hong, J. Li, Z. Jiang, X.-M. Ma, Y.-J. Hao, M.-Y. Zhu, W. Liu, M. Zeng, S. Jayaram, M. Lenger, J. Ding, S. Mo, K. Tanaka, M. Arita, Z. Liu, M. Ye, D. Shen, J. Wrachtrup, Y. Huang, R.-H. He, S. Qiao, Q. Liu, C. Liu, Observation of plaid-like spin splitting in a noncoplanar antiferromagnet. *Nature* **626**, 523–528 (2024).

28. E. Runge, E. K. U. Gross, Density-Functional Theory for Time-Dependent Systems. *Phys. Rev. Lett.* **52**, 997–1000 (1984).

29. G. Kresse, J. Furthmuller, Efficiency of ab-initio total energy calculations for metals and semiconductors using a plane-wave basis set. *Comput. Mater. Sci.* **6**, 15–50 (1996).

30. G. Kresse, J. Furthmuller, Efficient iterative schemes for ab initio total-energy calculations using a plane-wave basis set. *Phys. Rev. B* **54**, 11169–11186 (1996).

31. J. P. Perdew, K. Burke, M. Ernzerhof, Generalized Gradient Approximation Made Simple. *Phys. Rev. Lett.* **77**, 3865–3868 (1996).

32. G. Kresse, D. Joubert, From ultrasoft pseudopotentials to the projector augmented-wave method. *Phys. Rev. B* **59**, 1758–1775 (1999).

33. J. K. Dewhurst, S. Sharma, Elk code. *elk.sourceforge.net*.

Supplemental Materials for

**Ultrafast controlling net magnetization in g-wave altermagnets *via* laser fields**
Zhaobo Zhou et al.

**This PDF file includes:**
Spin-resolved LDOS of Cr1 and Cr2 atoms along the S′-A′-U′ path
Normalized spin moment of Cr1 and Cr2 with different $\theta$ and $\varphi$
Normalized spin moment of Cr1 and Cr2 with and without SOC
Schematic of $S_{VBM}$ in BZ with laser incidence at $\theta=30°$ $\varphi=-30°$ and $\theta=30°$ $\varphi=-90°$
Fig. S1 to S5

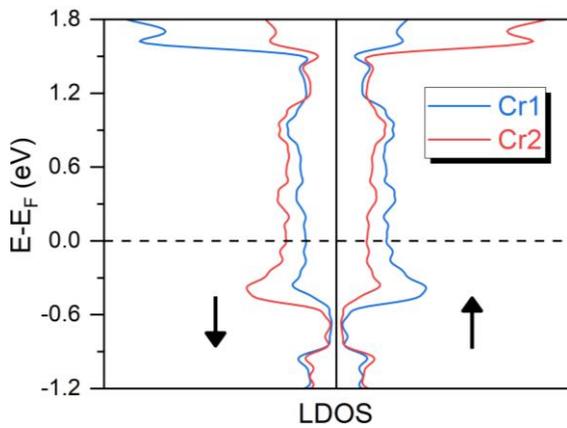

**Fig. S1.** Spin-resolved LDOS of Cr1 and Cr2 atoms along the S′-A′-U′ path.

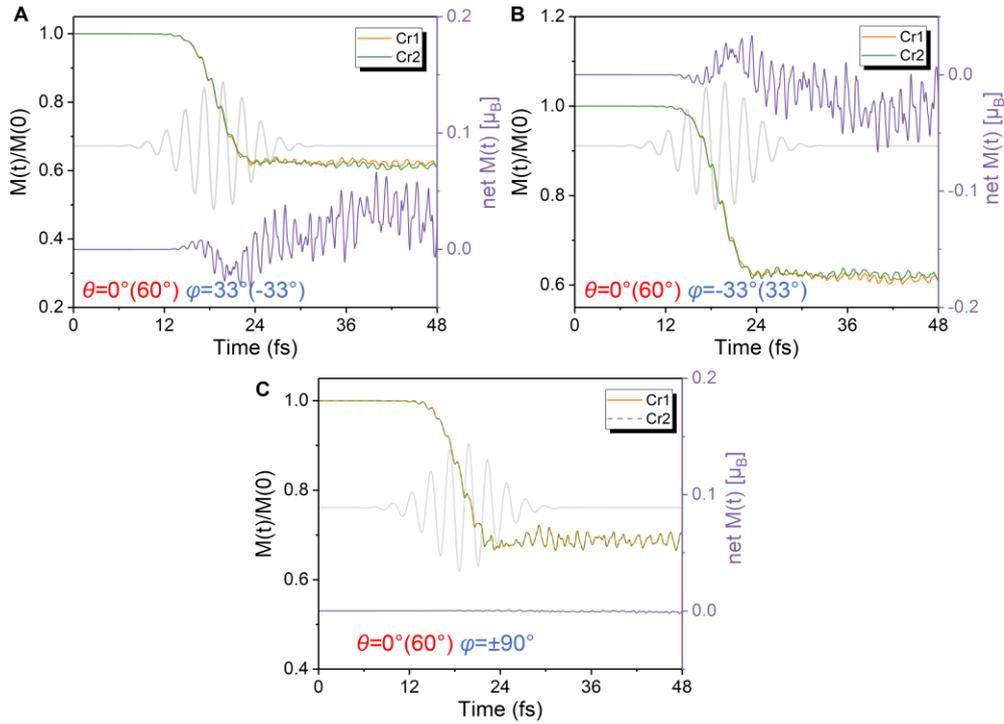

**Fig. S2.** Normalized atom-resolved spin moment of Cr1 and Cr2 as a function of time at (**A**) $\theta=0°$ (60°) $\varphi=33°$ (-33°), (**B**) $\theta=0°$ (60°) $\varphi=-33°$ (33°) and (**C**) $\theta=0°$ (60°) $\varphi=\pm90°$, respectively. The net magnetic moment is shown in purple.

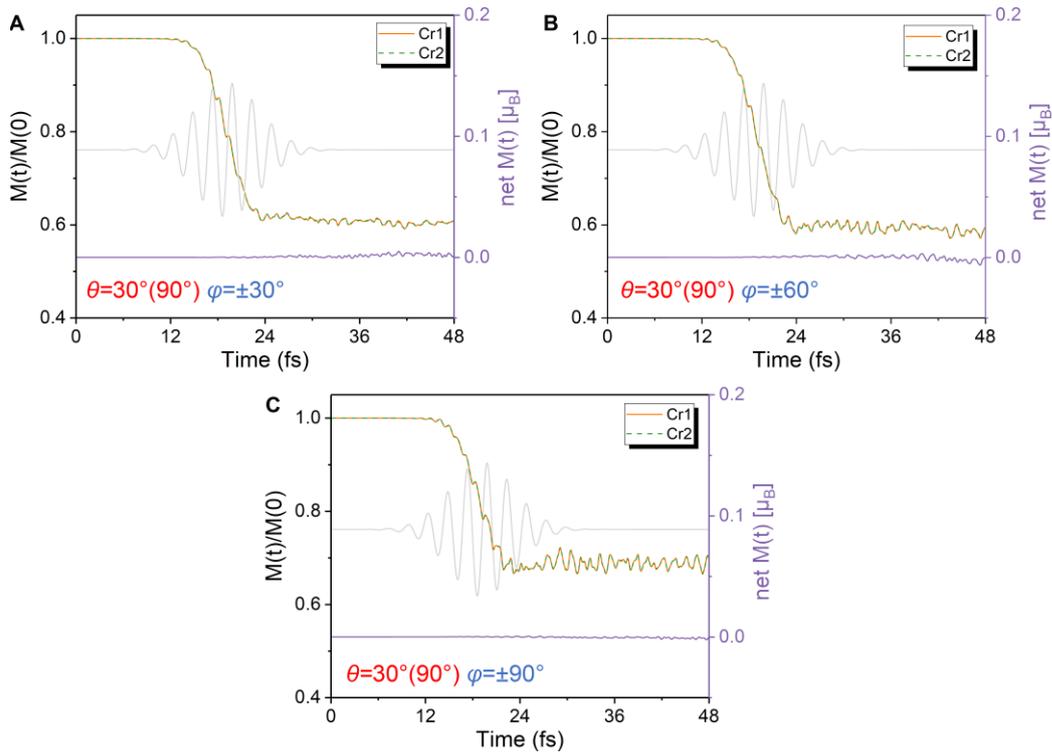

**Fig. S3.** Normalized atom-resolved spin moment of Cr1 and Cr2 as a function of time at $\theta=30°$ (90°) (**A**) $\varphi=\pm30°$, (**B**) $\varphi=\pm60°$ and (**C**) $\varphi=\pm90°$, respectively. The net magnetic moment is shown in purple.

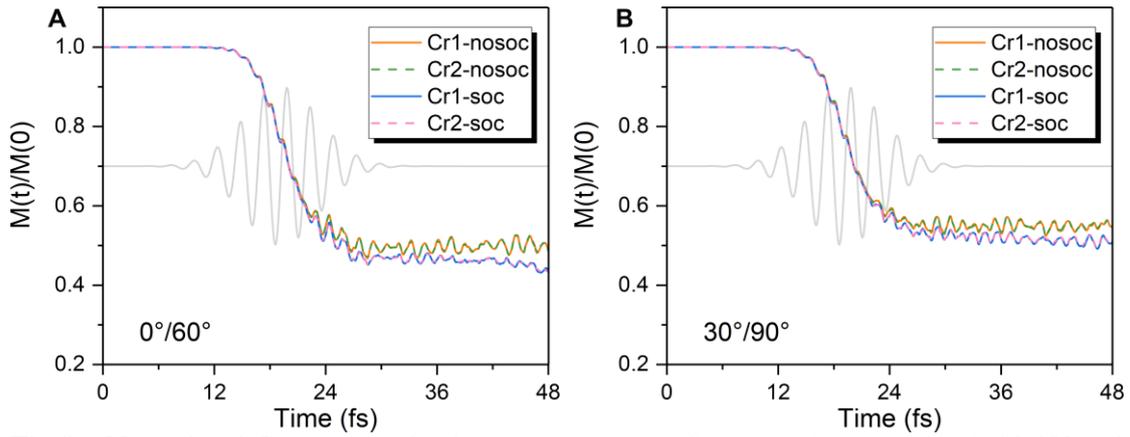

**Fig S4.** Normalized Cr atom-resolved spin moment as a function of time at (**A**) $\theta=0°/60°$ and (**B**) $\theta=30°/90°$ with and without SOC.

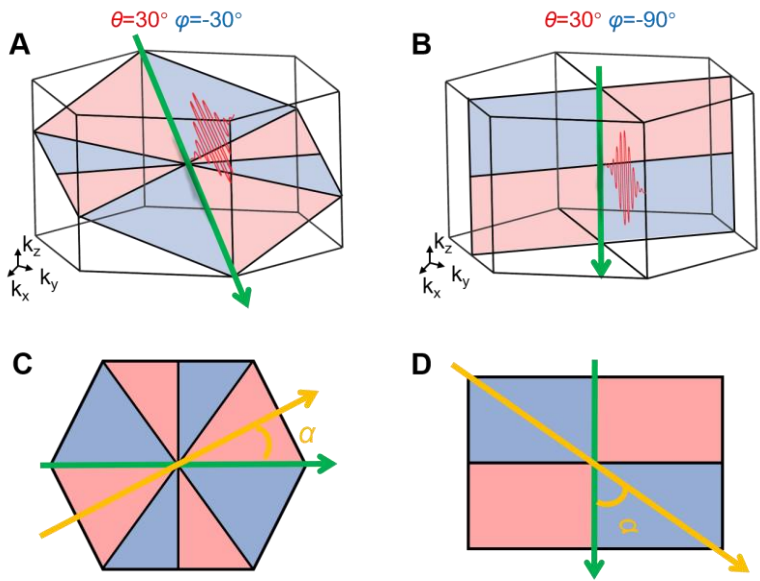

**Fig. S5.** (**A** and **B**) Schematic of $S_{VBM}$ in BZ with laser incidence at $\theta=30°$ $\varphi=-30°$ and $\theta=30°$ $\varphi=-90°$, respectively. (**C** and **D**) Corresponding schematic of 2D $S_{VBM}$. Green and yellow arrows represent the electric field vector of the laser with angle $\varphi$ (E$\varphi$) and the electric field vector after rotating the laser with an angle $\alpha$ (E$\alpha$).